\documentclass[twocolumn,preprintnumbers,amsmath,amssymb,superscriptaddress,prb]{revtex4}

\usepackage{graphicx,amsmath,amsfonts,amssymb}
\usepackage{dcolumn}
\usepackage{bm}
\usepackage{color}

\newcommand{\mbf}{\mathbf}

\begin{document}

\title{Strain-rate and temperature dependence of yield stress of amorphous solids via self-learning metabasin escape algorithm}

\author{Penghui Cao}
\author{Xi Lin\footnote{Corresponding author: linx@bu.edu}}
\author{Harold S. Park\footnote{Corresponding author: parkhs@bu.edu}}
    \affiliation{Department of Mechanical Engineering, Boston University, Boston, MA 02215, USA}


\begin{abstract}

A general self-learning metabasin escape (SLME) algorithm~\citep{caoPRE2012} is coupled in this work with continuous shear deformations to probe the yield stress as a function of strain rate and temperature for a binary Lennard-Jones (LJ) amorphous solid.  The approach is shown to match the results of classical molecular dynamics (MD) at high strain rates where the MD results are valid, but, importantly, is able to access experimental strain rates that are about ten orders of magnitude slower than MD.  In doing so, we find in agreement with previous experimental studies that a substantial decrease in yield stress is observed with decreasing strain rate.  At room temperature and laboratory strain rates, the activation volume associated with yield is found to contain about 10 LJ particles, while the yield stress is as sensitive to a $1.5\%T_{\rm g}$ increase in temperature as it is to a one order of magnitude decrease in strain rate.  Moreover, our SLME results suggest the SLME and extrapolated results from MD simulations follow distinctly different energetic pathways during the applied shear deformation at low temperatures and experimental strain rates, which implies that extrapolation of the governing deformation mechanisms from MD strain rates to experimental may not be valid. 


\end{abstract}



\maketitle


\section{Introduction}

The plasticity of amorphous solids such as metallic glasses has been extensively studied in recent years~\citep{schuhAM2007}.  A key parameter that must be accurately predicted in this regard is the yield stress, because amorphous solids typically fail catastrophically immediately following yield via shearbanding~\citep{chengPMS2011} due to their lack of strain hardening.   However, a definitive link between the effects of temperature and experimentally-relevant strain rates on the yield stress has not been established to-date.

There has recently been significant effort in studying the inelastic deformation of amorphous solids using atomistic simulation techniques such as classical molecular dynamics (MD)~\citep{falkPRE1998,caoAM2009,zinkPRB2006,shimizuAM2006,chengAM2011,shiAM2007,muraliPRL2011}.  However, MD simulations suffer from well-known issues related to strain rates that are about 10 orders of magnitude larger than experimentally accessible ones.  Other researchers have attempted to avoid the time scale and strain rate issues that are inherent to MD by utilizing athermal, quasistatic shear (AQS) of amorphous solids to study the mechanisms leading to strain localization~\citep{tsamadosPRE2009,maloneyPRE2006,tanguyEPJE2006,karmakarPRL2010,dasguptaPRL2012}.  Because there is no thermal energy in the system as would be the case in real experiments, the energetic barriers that are crossed on the potential energy surface (PES) are artificially low, and thus the system does not explore all possible configurations (i.e. deformation mechanisms) that it would experimentally.  A more recent approach to avoiding the limitations of MD to study the yielding and plasticity of amorphous solids is so-called PES exploration techniques~\citep{rodneyPRL2009,rodneyPRB2009,deloguPRL2008b,mayrPRL2006}.  These methods have had some success in calculating the activation energy and volumes of shear transformation zones (STZs)~\citep{deloguPRL2008b,mayrPRL2006}.  However, none of these studies has been able to explore a sufficiently large portion of the PES to make definitive statements about the strain rate and temperature-dependence of the yield stress for amorphous solids, particularly at laboratory strain rates.

The above discussion makes clear that there is a pressing need for advanced atomistic simulation techniques that are able to access experimental strain rates, and thus give new insights into the mechanical behavior and properties of amorphous solids at these slower strain rates.  Therefore, there are two key objectives of this work.  The first is to introduce a new approach to studying the mechanics of amorphous solids at strain rates ranging from experimental to those seen in MD simulations.  We accomplish this by presenting a new computational technique that couples the recently developed self-learning metabasin escape (SLME) algorithm for PES exploration~\citep{caoPRE2012,caoPRE2013} with shear deformation and the standard Monte Carlo approach.  As a first step, we verify the ability of the SLME approach to reproduce benchmark classical MD simulation results at high strain rates.  The second objective of this work is to determine the effects of strain rate and temperature on the yield stress of amorphous solids, with an emphasis on studying these quantities using the SLME method at experimentally relevant strain rates that MD cannot access.  The findings have implications for the validity of interpreting recently developed universal scaling laws for amorphous solids~\citep{johnsonPRL2005} within the context of extrapolating the results of high strain rate MD simulations to experimental strain rates, as has recently been proposed~\citep{chengAM2011}, and also for the universality of the yield mechanism~\citep{johnsonPRL2005} for amorphous solids for all strain rates and temperatures.  We also establish the strain rate equivalent of temperature on the yield stress at experimentally-accessible strain rates.

\section{Simulation Details}

\subsection{Binary Lennard-Jones Model}

A binary Lennard-Jones (bLJ) potential~\citep{kobPRE1995} is used in our simulations with periodic boundary conditions in all three directions.  This bLJ potential is widely utilized for atomistic studies of amorphous solids because its ground state is not crystalline.  The system contained $N= 1000$ particles of the same unit mass, with an A:B particle ratio of 4:1.  Standard parameters are used: $\epsilon_{\rm AA} = 1.0$, $\epsilon_{\rm AB} = 1.5$, $\epsilon_{\rm BB} = 0.5$, $\sigma_{\rm AA} = 1.0$, $\sigma_{\rm AB} = 0.8$, and $\sigma_{\rm BB} = 0.88$, while the cutoff distances are set to $2.5\sigma_{\rm AA}$.  Supercooled liquids were prepared by equilibrating the system at temperature $T= 2.0$ followed by slow quenching to $T= 1.0$ using the {\it NVT} ensemble at a constant particle density of $1.2$.  These supercooled liquids were then further quenched at constant {\it NPT} to obtain stress-free glassy structures at different target temperatures, $T=0.05, 0.1, 0.15, 0.2, 0.3$ and $0.4$ near and below the glass transition temperature $T_{g}=0.37$~\citep{kushimaJCP2009}, where all cooling was done at a constant rate of $4\times10^{-6}$~\citep{sastryNATURE1998}.  All units in this section, as well as the rest of the manuscript, are given in reduced (dimensionless) LJ form.  

\subsection{Self-Learning Metabasin Escape Algorithm}

\begin{figure} \begin{center} 
\includegraphics[scale=0.15]{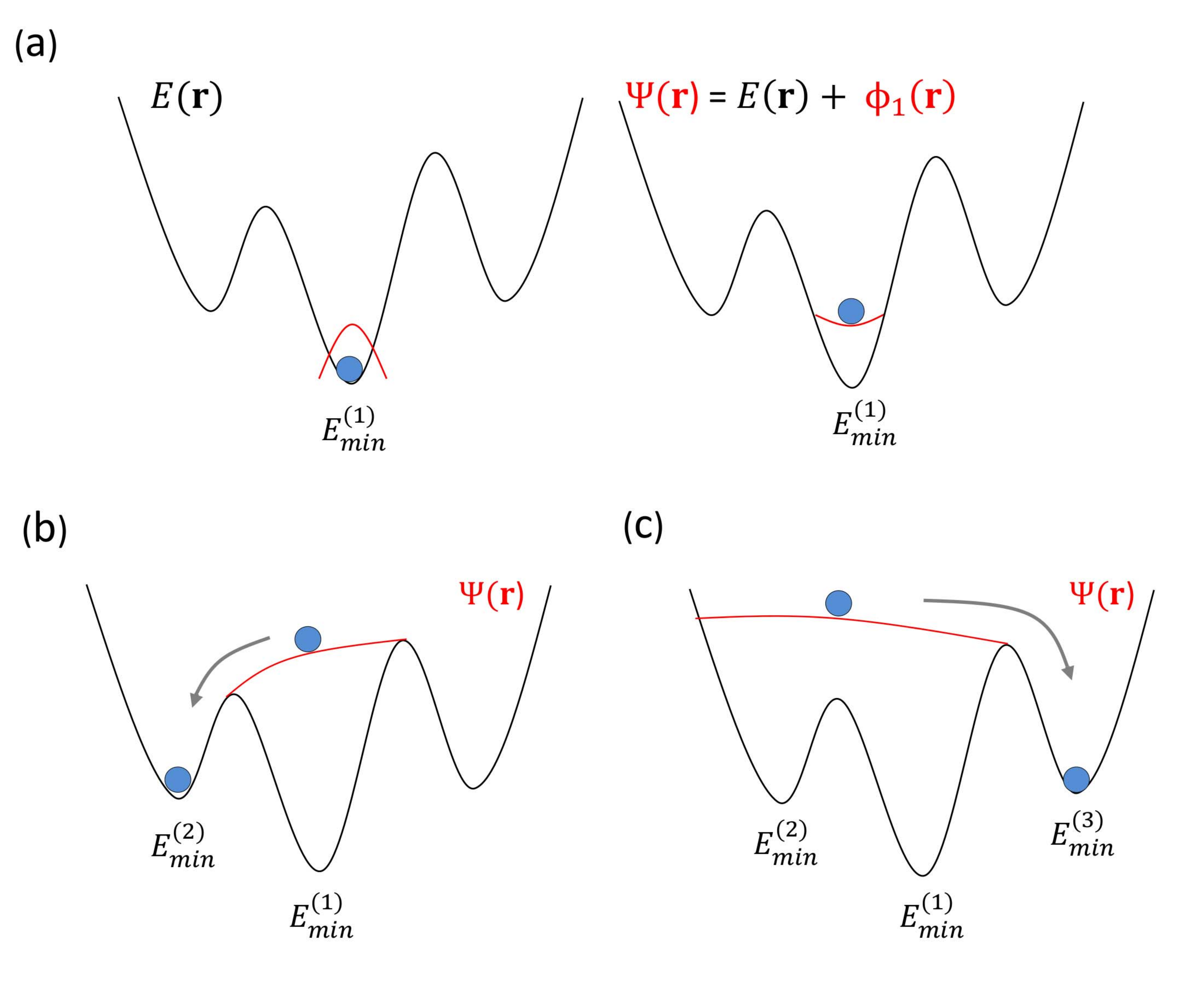}
\caption{\label{}Schematic of the autonomous basin climbing (ABC) potential energy surface exploration technique.  (a) Depiction of how the addition of a penalty function $\phi_{1}$ to the PES defined by $E(\mbf{r})$ results in the penalty function modified PES defined by $\Psi(\mbf{r})$.  (b) and (c) depict schematically how the addition of more penalty functions result in the system being pushed out via $\Psi(\mbf{r})$ of various local minima into other, neighboring energy basins.  $E_{min}$ and $E_{sad}$ correspond to energy minima and saddle points, respectively.  We emphasize that while the PES depicted in this figure is one-dimensional, the SLME algorithm utilized in this work explores the 3$N$-dimensional PES, where $N$ is the number of atoms in the simulation.}
\label{abc} \end{center} \end{figure}

We now overview the computational technique, the self-learning metabasin escape (SLME) algorithm that we use to explore the PES for each state of strain for a given strain rate.  As illustrated in Fig. \ref{abc},~\citet{kushimaJCP2009} recently developed the autonomous basin climbing (ABC) algorithm to explore the PES for a given atomistic system.  The ABC method works in a very intuitive manner.  Starting from any local energy minimum, penalty functions $\phi_{i}(\mbf{r})$ are successively applied in order to climb out of the current local minima well and explore other, neighboring energy wells.  Mathematically, this is written as
\begin{equation}\label{eq:abc1} \Psi(\mbf{r})=E(\mbf{r})+\sum_{i=1}^{p}\phi_{i}(\mbf{r}),
\end{equation}
where $\Psi(\mbf{r})$ is the augmented potential energy due to the addition of the penalty functions, $E(\mbf{r})$ is the original potential energy function, i.e. the bLJ potential in the present case, and $p$ is the total number of penalty functions.  Although in principle any type of localized functions (i.e. Gaussians~\citep{laioPNAS2002,kushimaJCP2009,kushimaJCP2009_2}) can be used in Eq. (\ref{eq:abc1}), we chose quartic penalty functions in this work due to their desirable property of naturally vanishing energy and forces at the penalized subspace boundaries~\citep{caoPRE2012}.

As can be inferred from Eq. \ref{eq:abc1}, many small penalty functions are needed in order to push the system out of a given energy basin.  However, all of these penalty functions must be kept such that the system does not fall back into an energy basin that has already been explored.  Clearly, the requirement to store all previous penalty functions becomes expensive as more and more energy basins are explored.  Because of this, the computational expense associated with the ABC method increases substantially, and becomes the bottleneck of the ABC method, as the PES exploration continues.  

This issue was alleviated substantially in the recent work of~\citet{caoPRE2012}, where a novel, self-learning combination scheme was implemented.  Specifically, self-generated and self-reconstructed penalty functions were imposed to assist the system in escaping from a given local energy basin.  After each independent penalty function is added, a search for the local energy minimum is performed on the augmented (i.e. $\Psi(\mbf{r})$ in Eq. (\ref{eq:abc1}), or as illustrated in Figure (\ref{abc})) energy surface, which is the sum of the original potential energy $E(\mbf{r})$ and all the previously imposed penalty potentials $\phi_{p}(\mbf{r})$.  The essential idea is that instead of storing all of the (many) penalty functions that have been used to boost the system out of the different energy wells it has explored, the penalty functions are combined in various ways such that, upon exiting a given energy well, only a few penalty functions remain.  This approach, called the SLME approach, and the resulting decrease in penalty function storage requirements, was shown to lead to an exponential increase in computational efficiency as compared to the previous ABC implementation~\citep{caoPRE2012}.  By repeating the alternating sequence of penalty function addition and augmented energy relaxation, the system is self-activated to fill up the local energy basin and escape through the lowest saddle point.  By maintaining all the independent penalty functions imposed during the SLME trajectory, frequent recrossing of small barriers is eliminated, which is a significant advantage of such history-penalized methods~\citep{laioPNAS2002,kushimaJCP2009,caoPRE2012}.  We emphasize that while Fig. \ref{abc} depicts the ABC method in 1D, in actuality for the present work the SLME approach investigates the entire, 3$N$-dimensional (3$N$-D) PES, where $N$ is the total number of atoms in the system.

We note the benefits of using the SLME method as compared to other PES exploration techniques.  For example, both the hyperdynamics~\citep{voterPRL1997} and dimer methods~\citep{henkelmanJCP1999} rely on penalizing along the softest eigenmode direction within a given energy basin which is not necessarily aligned with the true activation eigenmode.  More seriously, without keeping the penalty functions in those previously visited PES energy wells, the system generally spends most of its computational time re-visiting frequent but nonessential events.  For both the metadynamics~\citep{laioPNAS2002} and bond-boost accelerated MD~\citep{mirona03}, one needs to specify a small set of order parameters based on chemical intuition, and then restrict the activation search only inside this order-parameter subspace. It is also relevant to discuss this approach in contrast to nudged elastic band (NEB) techniques that have recently been utilized to study the deformation mechanisms of nanostructured metals at experimentally-relevant time scales~\citep{zhuPRL2008}.  The NEB approach is particularly well-suited for metal plasticity because it requires {\it a priori} knowledge of the final configuration in order to find the minimum energy pathway.  In the case of metals, it is well-known that crystal defects such as twins, dislocations and stacking faults are the likely plastic deformation mechanisms~\citep{parkJMPS2006}.  However, the situation is quite different for amorphous solids, where the atomic structure of the equivalent basic deformation mechanism, the STZ, remains unknown~\citep{schuhAM2007,chengPMS2011}.  

\begin{figure*} \begin{center} 
\includegraphics[scale=0.3]{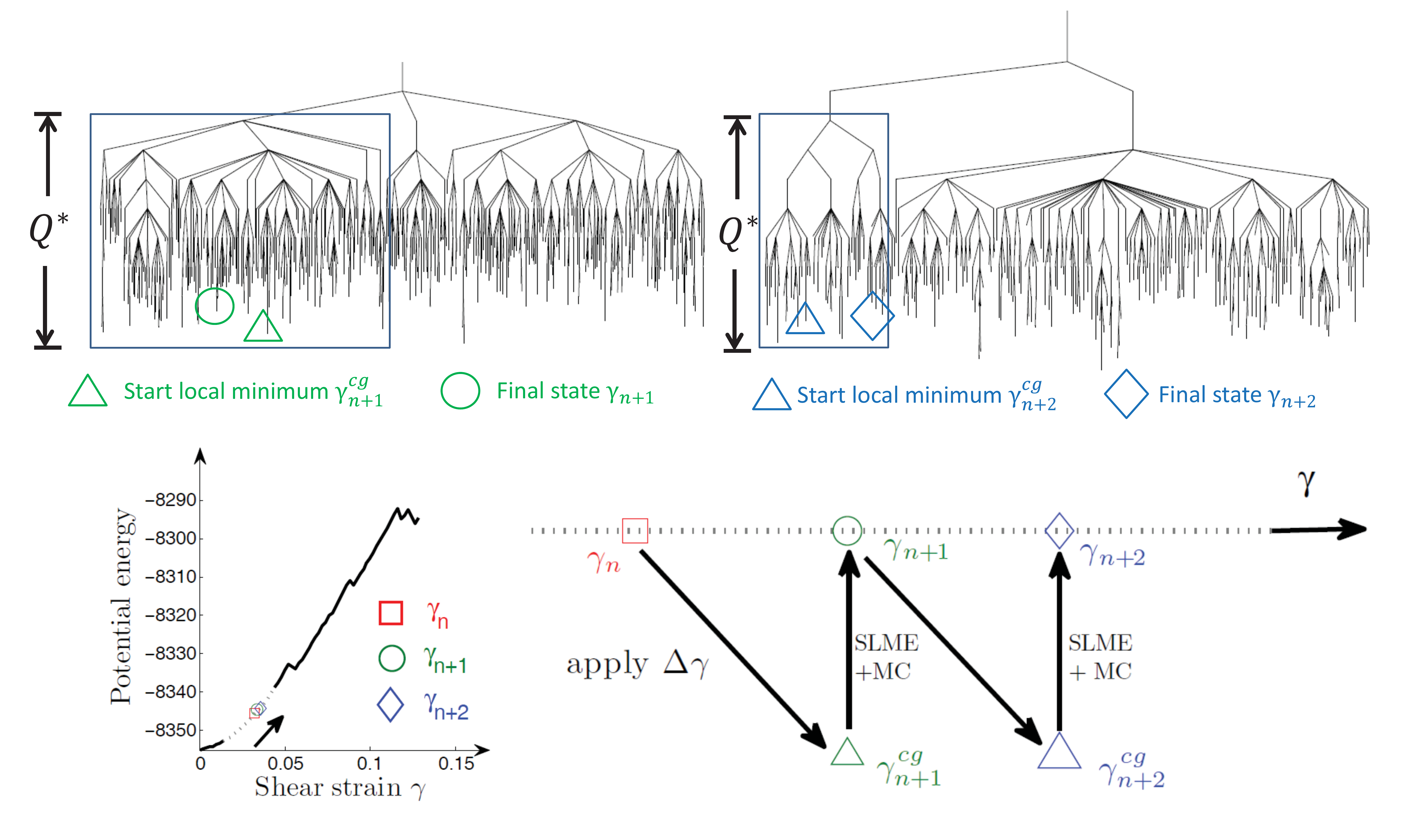}
\caption{\label{}Illustration of how the SLME method~\citep{caoPRE2012} is utilized to find the equilibrium configuration after a given strain increment $\Delta\gamma$ is applied to the system.  Specifically, starting from $\gamma_{n}$, a shear increment $\Delta\gamma$ is applied to the system.  At that point, a standard conjugate gradient (CG) energy minimization is performed while keeping the strain fixed, giving the state $\gamma_{n+1}^{cg}$.  Starting from the energy minimized configuration $\gamma_{n+1}^{cg}$, the SLME method is used to determine the potential energy tree structure as shown, where the lower end point of each vertical line specifies an independent local minimum energy configuration, and where every pair of these local minima are connected by a unique saddle point specifying the lowest activation energy barrier between them.  We truncate the tree structure to only enable energy transitions below $Q^{*}$, as shown in the blue box.  Finally, a classical monte carlo algorithm is employed to find, amongst the hundreds of local minima in the blue box, the most likely equilibrium configuration, which is then denoted $\gamma_{n+1}$. The same procedure is then utilized to find the next equilibrium configuration for the strain $\gamma_{n+2}$, though we note that the PES tree structure is different at the new shear strain $\gamma_{n+2}$, which again is mapped out using the SLME method.}
\label{fig1} \end{center} \end{figure*}

The SLME approach thus resolves two critical issues with regards to mapping out the PES of amorphous solids.  First, the SLME method enables, along with very few other computational techniques~\citep{mironPRL2004}, one to systematically explore sequential metabasin activation events on the complete 3$N$-dimensional (3$N$-D) PES without {\it a priori} knowledge of final states or order parameters as required for supercooled liquids and amorphous solids.  Second, the computational efficiency of the SLME approach is critical as it enables us to get access to a sufficiently large configurational space by infrequent free energy activation events over very large activaton free energy barriers $Q^{*}(T,\dot{\gamma})$ that are needed to access temperatures $T$ well below the glass transition temperature and at the laboratory shear strain rates $\dot{\gamma}$, as to be discussed in the next section.  However, the SLME approach is just a PES exploration technique, and does not explain how we incorporate temperature and strain rate effects.  We explain how this is done in the following section.
 
\subsection{Shear-Coupled Self-Learning Metabasin Escape Algorithm}

\subsubsection{Temperature and strain-rate dependent activation free energy $Q^*(T,\dot{\gamma})$ formalism}

To incorporate the effects of strain rate and temperature, we begin with the following expression for a single-event shear strain rate, which was derived by~\citet{zhuPRL2008} from transition state theory (TST) assuming constant temperature and strain rate:
\begin{equation}\label{eq:tst3}
\dot{\gamma}_{\rm single}=nv_0\dfrac{k_{\rm B} T}{\mu \Omega}\exp\left[-\dfrac{Q(T)-TS_{\rm c}}{k_{\rm B} T}\right],
\end{equation}
where $n$ is the number of independent nucleation sites, $v_0$ is the attempt frequency, $\mu$ is the shear modulus, $\Omega$ is the activation volume, and $S_{\rm c}$ is the activation configurational entropy, which has previously been calculated for crystalline FCC metals for the specific case of dislocation nucleation~\citep{ryuPNAS2011}.

It is important to note that the single-event activation energy $Q(T)$ in Eq. \ref{eq:tst3} is explicitly dependent on the temperature $T$, obtained naturally from the basin-filling trajectories ~\citep{liPO2011,kushimaJCP2009}.  Following this formalism~\citep{liPO2011,kushimaJCP2009}, we now describe how to extend the \emph{single-event} activation free energy $Q(T)$ to the temperature and strain-rate dependent \emph{many-event} $Q^*(T,\dot{\gamma})$, where $Q^{*}(T,\dot{\gamma})$ contains many (hundreds) of such activation events, as illustrated by the blue box in Fig. \ref{fig1}.  Specifically, $Q^{*}(T,\dot{\gamma})$  is the maximal activation energy with respect to the initial free-energy basin $F(T)$, so that $Q^*(T,\dot{\gamma})$ truncates the ergodic Markovian system into an ergodic Markovian subspace and the remainder, the part that is not accessible at the given strain rate $\dot{\gamma}$.  As $\dot{\gamma}$ decreases, the ergodic Markovian subspace increases monotonically, with the important implication that more and more mechanical deformation pathways that were not accessible at high strain rates become accessible assuming that the PES exploration technique (i.e. the SLME approach) is able to reach and climb over the corresponding energy barriers on the PES.  Because the SLME approach enables us to efficiently access and calculate the allowed activated states $Q(T,\dot{\gamma}) \le Q^*(T,\dot{\gamma})$ for essentially arbitrarily large $Q^{*}(T,\dot{\gamma})$, we are able to compute the yield stress $\tau (T, \dot{\gamma})$ and activation volume $\Omega(T,\dot{\gamma})$ at all relevant temperatures and shear strain rates $\dot{\gamma}$ ranging from MD to experimentally-accessible.

Having established the theoretical basis for extending $Q(T)$ to $Q^{*}(T,\dot{\gamma})$, we note that in contrast to the simpler deformation processes occurring in crystalline materials~\citep{zhuPRL2008}, the coupled thermomechanical deformation events in amorphous solids likely consist of multiple sequential activation events.  

Therefore, by defining a characteristic temperature-dependent prefactor $\dot{\gamma_0}(T)= \dfrac{k_{\rm B} Tnv_0}{\mu \Omega}\exp\left(\dfrac{S_{\rm c}}{k_{\rm B}}\right)$\citep{johnsonPRL2005,chengAM2011}, we can rewrite Eq. (\ref{eq:tst3}) as
\begin{equation}\label{eq:tst3a}
\dot{\gamma}_{\rm single}=\dot{\gamma_{0}}\exp\left[-\dfrac{Q(T)}{k_{\rm B} T}\right].
\end{equation}
Finally, by converting from $Q(T)$ to $Q^{*}(T,\dot{\gamma})$ based on the above discussion, we can construct the maximal activation energy barrier by rearranging Eq. (\ref{eq:tst3a}) as
\begin{eqnarray}\label{eq:tst4}
Q^*(T,\dot{\gamma})= -k_{\rm B} T \ln\left(\dfrac{\dot{\gamma}}{\dot{\gamma_{0}}}\right),
\end{eqnarray}
which defines the ergodic Markovian region in the entire SLME connectivity tree structures at the given strain rate, for example as shown as the green box in Fig. \ref{fig1a}.  Within this ergodic window, all the transitions follow the Markov chain processes to reach a local equilibrium, so that the amorphous solid (bLJ) system can relax to the accessible lowest free energy configurations.  

In summary, Eq. (\ref{eq:tst4}) plays the essential role of being a physical link between the SLME trajectories and the strain rates from MD to experimental values, depending on only a single unknown temperature-dependent prefactor $\dot{\gamma_0}(T)$.  It is important to note that variations in the defined strain rates $\dot{\gamma}$ of Eq. (\ref{eq:tst3a}) are dominated by their exponential dependence on the activation energies $Q(T)$ within the truncated ergodic SLME energy window $Q^*(T,\dot{\gamma})$, and only slightly correlated to the temperature-dependent prefactor $\dot{\gamma_0}(T)$.  Therefore, this prefactor can be either simplified to be a temperature-independent constant, or determined by fitting to the results of high strain MD simulations. In the present work, we obtain the prefactor by fitting to the MD results as discussed in Section 3.1.

\subsubsection{Algorithmic Details}

We now detail how the SLME method is coupled with shear deformation and classical monte carlo to calculate the stress and equilibrium atomic positions of the bLJ solid as a function of strain, strain rate and temperature, with no change in methodology needed to distinguish between elastic and plastic strain increments.  After obtaining the initial stress-free glassy structures for a given temperature, we apply the following algorithm for all loading increments.  Specifically, assume that as shown in Fig. \ref{fig1}, the system exists at shear strain $\gamma_{n}$.  We then apply a shear strain increment $\Delta\gamma$=0.1\%, followed by a standard conjugate gradient energy minimization to find the resulting equilibrium positions of the atoms, which brings us to the shear strain state $\gamma_{n+1}^{cg}$ in Fig. \ref{fig1}.  It is important to note that the system size and boundaries are held fixed during the energy minimization such that the shear strain $\gamma_{n+1}^{cg}=\gamma_{n}+\Delta\gamma$.

From that point, the SLME approach~\citep{caoPRE2012,caoPRE2013,kushimaJCP2009} is utilized to explore the PES at the strain $\gamma_{n+1}^{cg}$, as illustrated via the potential energy connectivity tree structures~\citep{wales2003} shown in Fig. \ref{fig1}, while again the system size and boundaries is held fixed.   Importantly, we only allow transitions within the SLME connectivity tree structures below the maximum energy barrier $Q^{*}$ shown in Eq. (\ref{eq:tst4}) as highlighted by the blue box shown in Fig. \ref{fig1}.  The maximum energy barrier $Q^{*}$ is a defined parameter that specifies the maximum barrier height on the PES that can be overcome, via thermal assistance, for a given strain rate.  This is because in physical terms, choosing a value of $Q^{*}$ is equivalent, as shown in Eq. (\ref{eq:tst4}), to specifying the strain rate of the simulation for a given temperature.  In other words, for very high strain rates as seen in MD simulations, only small energetic barriers $Q^{*}$ can be crossed for each strain increment due to the small amount of time given to the system to explore other possible, thermally-assisted configurations.  In contrast, at slower strain rates, the system has more time between successive strain increments such that it can sufficiently explore many other possible, thermally-assisted configurations, and thus potentially climb over larger energy barriers, with the sole restriction that the thermally assisted barrier crossing must be smaller than $Q^{*}$.  It is important to note, however, that we do not enforce that the maximum barrier height $Q^{*}$ is crossed for each strain increment.

Summarized a different way, the picture of deformation underlying our work is one that receives contributions due to both mechanical and thermal work.  The mechanical work dominates the deformation process at high strain rates, when the time in between strain increments is not sufficient to enable substantial, thermally-assisted atomic motion.  Thermal work is viewed as making a substantial contribution to the deformation process at slower strain rates, when sufficient time to enable thermally-driven deformation in between successive strain increments is provided to the system.

As shown in Fig. \ref{fig1} starting from the specific strain state $\gamma_{n+1}^{cg}$, the SLME algorithm typically finds on the order of a few hundred local mimina for each value of shear strain, which gives on the order of ten thousand local minima for the entire shear deformation process, as well as all of the corresponding lowest energy barriers between every pair of these local minima.  In other words, at a given strain rate, the system can self-explore the PES via the SLME approach by climbing over all the allowed energy barriers that are smaller than $Q^{*}$, as depicted via the blue boxed portion of the PES connectivity tree structure in Fig. \ref{fig1}.  Within this truncated potential energy subspace, we identify the most likely free energy basin, namely the basin with the lowest free energy at this instantaneous $(NVT)$-ensemble at the given strain state, via the standard Monte Carlo method.  This lowest free energy basin at strain $\gamma_{n+1}=\gamma_{n+1}^{cg}=\gamma_{n}+\Delta\gamma$, as shown by the green circle in Fig. \ref{fig1}, is assigned to be the initial configuration for the next loading increment.  Furthermore, the atomic configuration corresponding to the lowest free energy basin corresponds to the shear strain state $\gamma_{n+1}$.  The shear stress corresponding to the shear strain $\gamma_{n+1}$ is then obtained by calculating the virial stress based upon the atomistic configuration at $\gamma_{n+1}$.  At this point, a new shear strain increment of 0.1\% is applied and the SLME process as just described is repeated until the yield stress is obtained, where the yield stress is determined to be the maximum stress that is reached before the first substantial stress drop signifying yield is obtained.  We note that we also tested smaller strain increments of 0.01\%, which did not impact the value of the yield stress we report.

\section{Numerical Results and Discussion}

\subsection{Benchmarking SLME Results Against Classical MD at High Strain Rates}

\begin{figure} \begin{center}
\includegraphics[scale=0.65]{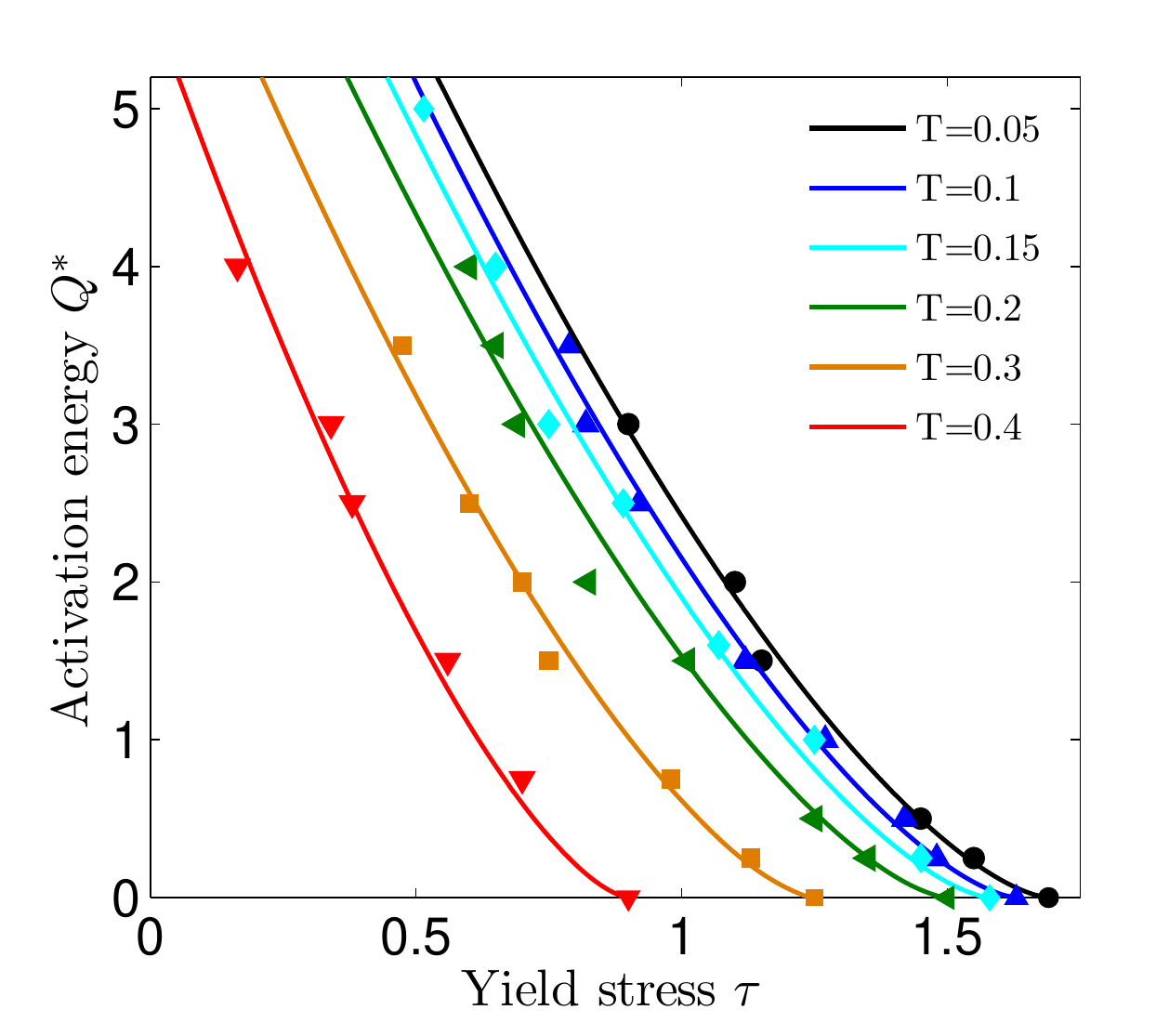} 
\caption{\label{}Shear yield stress $\tau$ as a function of activation energy $Q^{*}$ computed by the SLME method (filled symbols) at various temperatures $T$.  By fitting to Eq. (\ref{eq:qatau}) (solid lines), $A_T/(k_{\rm B}T)$ is found to be 185.4, 89.5, 57, 39.8, 22.83 and 14.25 for $T$=0.05, 0.10, 0.15, 0.20, 0.30, and 0.40, respectively.}
\label{fig2} \end{center} \end{figure}

Before discussing the key results of this manuscript, we first compare the SLME results to those obtained using high strain rate classical MD. This is done to ensure that the SLME method is able to reproduce the high strain rate MD results, which serve as the benchmark solution for the high strain rate regime.  Fig. \ref{fig2} shows the raw data output (filled symbols) from the SLME trajectories, specifically, the shear yield stress $\tau$ as a function of the ergodic activation energy window $Q^{*}$ for temperatures ranging from just above the glass transition temperature $T_{\rm g}= 0.37$~\citep{kushimaJCP2009} to deeply below. There are no error bars in Fig. \ref{fig2} as it is based on a single set of calculations starting from the same initial configuration for each temperature.  It is clear from Fig. \ref{fig2} that the yield stress in shear for a given temperature increases with increasing strain rate (equivalently with a decrease in the accessible activation energy window $Q^{*}$), and decreases with increasing temperature.  

Following~\citet{johnsonPRL2005} to assume the elastic energy density as a function of strain varies sinusoidally on average for amorphous solids, we fit the SLME data in Figure (\ref{fig2}) with solid lines using the formula 
\begin{eqnarray}\label{eq:qatau}
Q^{*}(\dot{\gamma}, T) = A_T(1-\tau/\tau_0)^{3/2}, 
\end{eqnarray}
where $\tau_{0}$ is the SLME-computed yield stress at $Q^{*}=0$ and $A_{T}$ is a fitting constant. The fixed exponent 3/2 comes from~\citet{johnsonPRL2005}, who observed a universal $(T/T_{\rm g})^{2/3}$ scaling law for plastic yielding of 31 different metallic glasses as
\begin{eqnarray}\label{eq:qatau2}
\tau= \tau_0-\tau_0\left[\dfrac{k_{\rm B}T_{\rm g}}{A_T}\ln\left(\dfrac{\dot{\gamma_0}}{\dot{\gamma}}\right)\right]^{2/3}\left(\dfrac{T}{T_{\rm g}}\right)^{2/3}. 
\end{eqnarray}
In Eq. (\ref{eq:qatau2}), there are two unknown temperature-dependent parameters to be determined based on the SLME results.  First, the effective activation barrier at the zero stress state, $A_{T}$ can be obtained from fitting the raw SLME data in Fig. \ref{fig2} to Eq. (\ref{eq:qatau}).  Before moving forward, we discuss the implications of the strong temperature-dependence for the fitting constant $A_{T}$ in Fig. \ref{fig2}, and its relationship to Eq. (\ref{eq:qatau}).  Our SLME results in Fig. \ref{fig2} demonstrate that the 3/2 exponential function in Eq. (\ref{eq:qatau}), which has become widely utilized since the work of~\citet{johnsonPRL2005}, can be used to fit the yield stress of amorphous solids for strain rates ranging from exceptionally fast (i.e. MD time scales) to experimentally-relevant.  However, in order to be applicable for the wide range of temperatures from far below $T_{g}$ to $T_{g}$ as in Fig. \ref{fig2}, $A_{T}$ must vary as a function of temperature.  This can be observed by the fact that the normalized value $A_{T}/(k_{\rm B}T)$ must change by more than an order of magnitude from 185.4 at $0.14T_{g}$ to 14.25 at $T_{g}$ in order to fit the different temperature-dependent curves in Fig. \ref{fig2}.  This also illustrates the error in assuming $A_{T}$ is constant, as has been done in previous works for amorphous solids~\citep{chengAM2011}.

\begin{figure*} \begin{center} 
\includegraphics[scale=0.67]{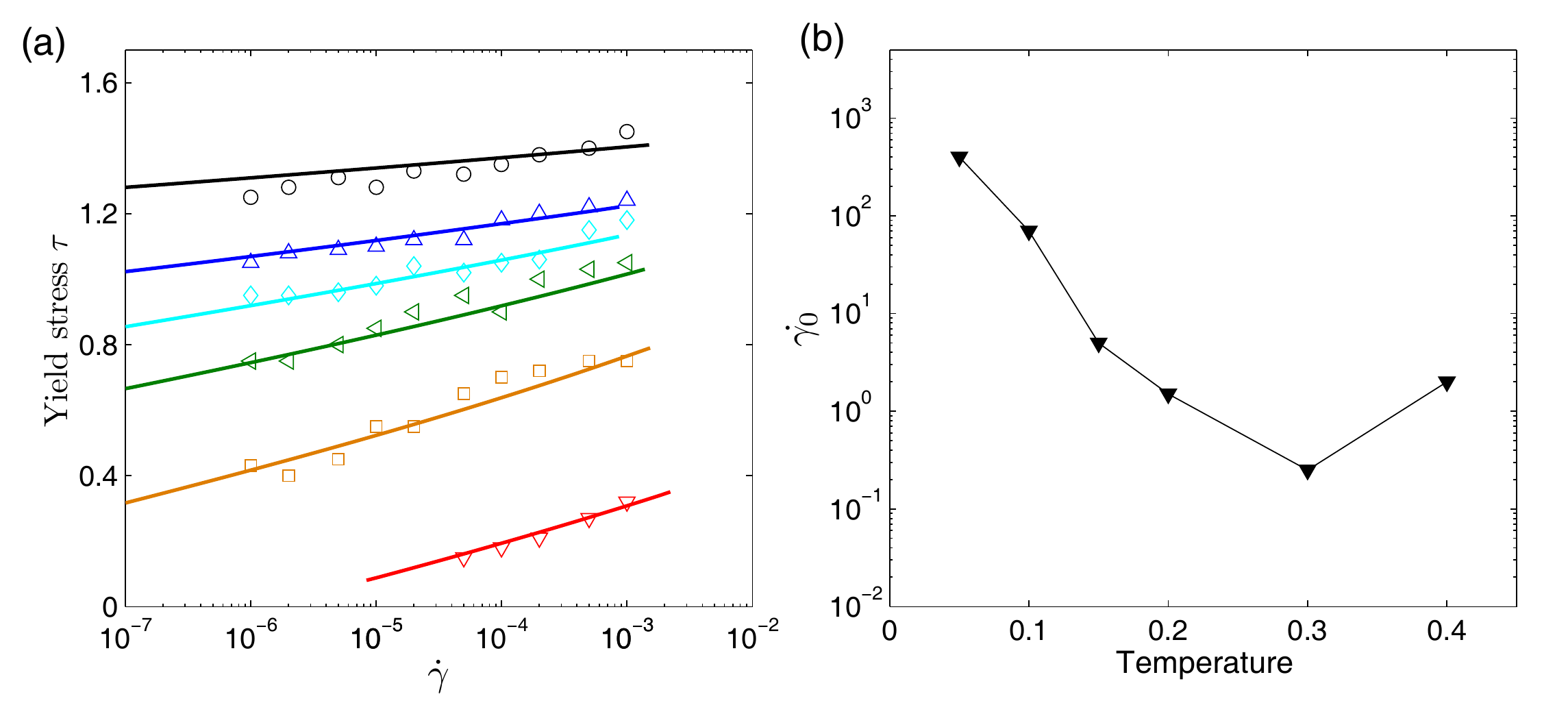}
\caption{\label{}(a) Shear yield stress  $\tau$ computed using the SLME (solid lines) and MD (open symbols) methods at the MD-accessible strain rates $\dot{\gamma}$.  (b) Values of the prefactor $\dot{\gamma_{0}}$ as a function of temperature.}
\label{fig1a} \end{center} \end{figure*}

The second more subtle parameter to be fit in Eq. (\ref{eq:qatau2}) is the temperature-dependent prefactor $\dot{\gamma_0}(T)=\dfrac{k_{\rm B} Tnv_0}{\mu \Omega}\exp\left(\dfrac{S_{\rm c}}{k_{\rm B}}\right)$, which was first defined above Eq. (\ref{eq:tst3a}).  This prefactor is difficult to calculate primarily due to issues in obtaining an exact value for the activation entropy $S_{c}$, but also because the variation of the attempt frequency $v_{0}$ with temperature is also unknown.  Furthermore, recent work has demonstrated that the attempt frequency $v_{0}$ can also take on a wide range of values~\citep{koziatekPRB2013,rodneyMSMSE2011}.  Therefore, in the present work, we obtain $\dot{\gamma_{0}}(T)$ by directly fitting to MD simulation results at various temperatures.  In doing so, we demonstrate in Fig. \ref{fig1a}(a) that the SLME results for the yield stress do match the MD simulation results for a range of temperatures for all accessible MD strain rates.  From Fig. \ref{fig1a}(b), it is clear that $\dot{\gamma_{0}}(T)$ increases significantly for temperatures below about T=0.2, while taking values close to 1 for larger temperatures.  This is likely due to the exponential dependence of the activation entropy $\exp\left(\dfrac{S_{\rm c}}{k_{\rm B}}\right)$ in low-temperature activation processes, i.e., a substantial activation entropy increase is needed for any activation events to occur at these low temperatures.

We make one other important comment regarding the fitting of $\dot{\gamma_{0}}(T)$.  Specifically, while we have obtained it for the SLME simulations for each temperature from MD simulations, the only impact of $\dot{\gamma_{0}}(T)$ on the SLME results in Fig. \ref{fig1a}, and also to be shown later in Fig. \ref{fig4} is to shift the position of the SLME curves, while importantly the slope remains unchanged.  Therefore, even though the SLME results are fit to match the MD results at MD strain rates, we will show later in Fig. \ref{fig4} because the slope of the SLME and MD curves differ even at high strain rates where the differences in the yield stress values are nearly negligible, the different slopes will cause noticeable differences between the SLME and MD results at experimental strain rates, particularly at low temperatures.

\begin{figure*} \begin{center} 
\includegraphics[scale=0.8]{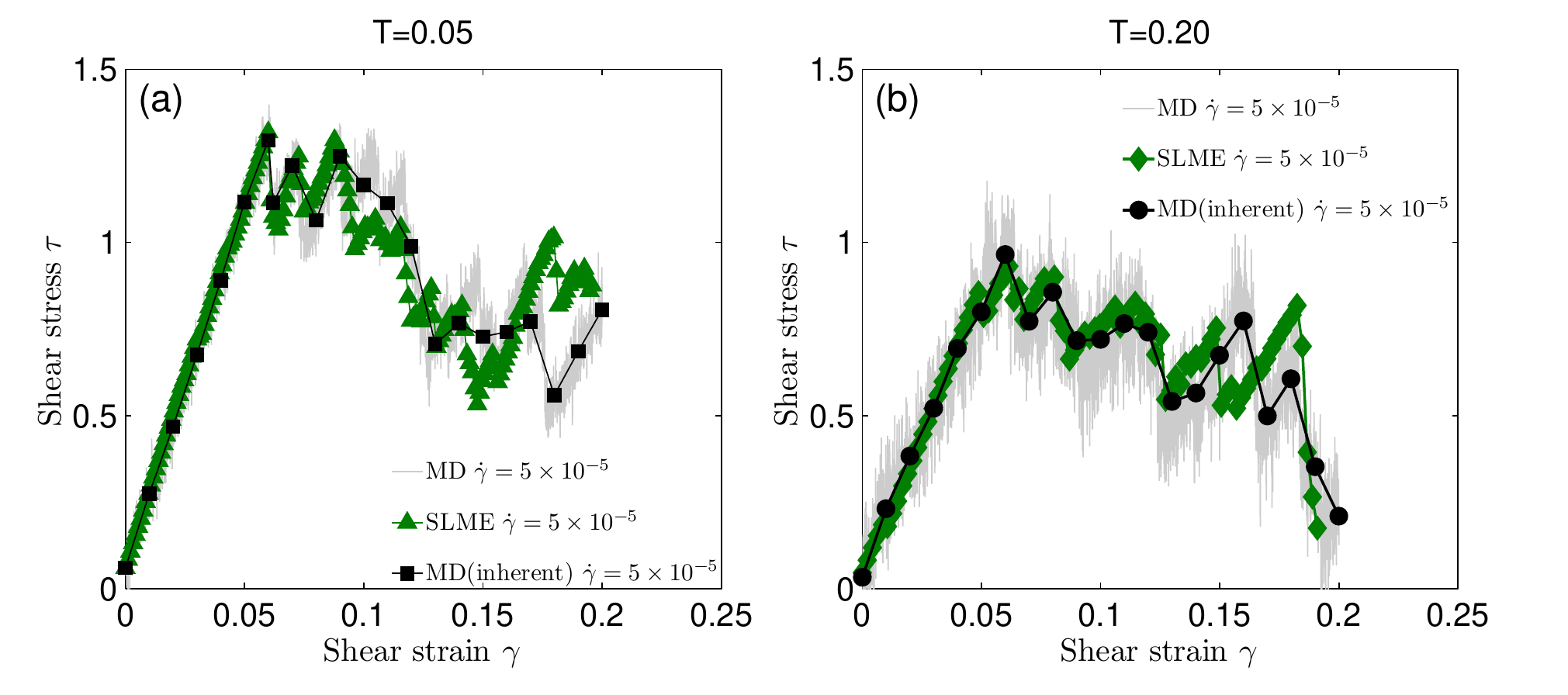}
\caption{\label{} Shear stress $\tau$ versus strain $\gamma$ at (a) $T=0.05$ and (b) $T=0.20$ by MD, MD (inherent) and SLME for MD-accessible strain rate of $\dot{\gamma}=5\times10^{-5}$.}
\label{fig13a} \end{center} \end{figure*}

As a final test of the SLME method to reproduce the benchmark MD simulation results at high, MD-accessible strain rates, we show in Fig. \ref{fig13a} a comparison of the SLME and MD-generated stress-strain curves for two different temperatures, $T=0.05$ and $T=0.2$, for the MD-accessible strain rate of $\dot{\gamma}=10^{-5}$.  As can be seen, there is more fluctuation in the MD results due to the dynamic nature of an MD simulation.  To smooth out the fluctuations to enable a clearer comparison between MD and SLME, we obtained the inherent structures for the MD simulations by performing energy minimization on the MD configuration to bring the MD trajectory to the nearest local minimum, which is shown by the solid black symbols in Fig. \ref{fig13a}.  Encouragingly, the SLME results agree with MD for both the low temperature case in Fig. \ref{fig13a}(a) where $T=0.05$, and the higher temperature case where $T/T_{g}=0.54$ in Fig. \ref{fig13a}(b), in both the yield stress and strain, as well as the general trajectory of the system after yielding has occurred.

\subsection{SLME Results of Strain Rate and Temperature-Dependence of Yield Stress}

\begin{figure} \begin{center}
\includegraphics[scale=0.8]{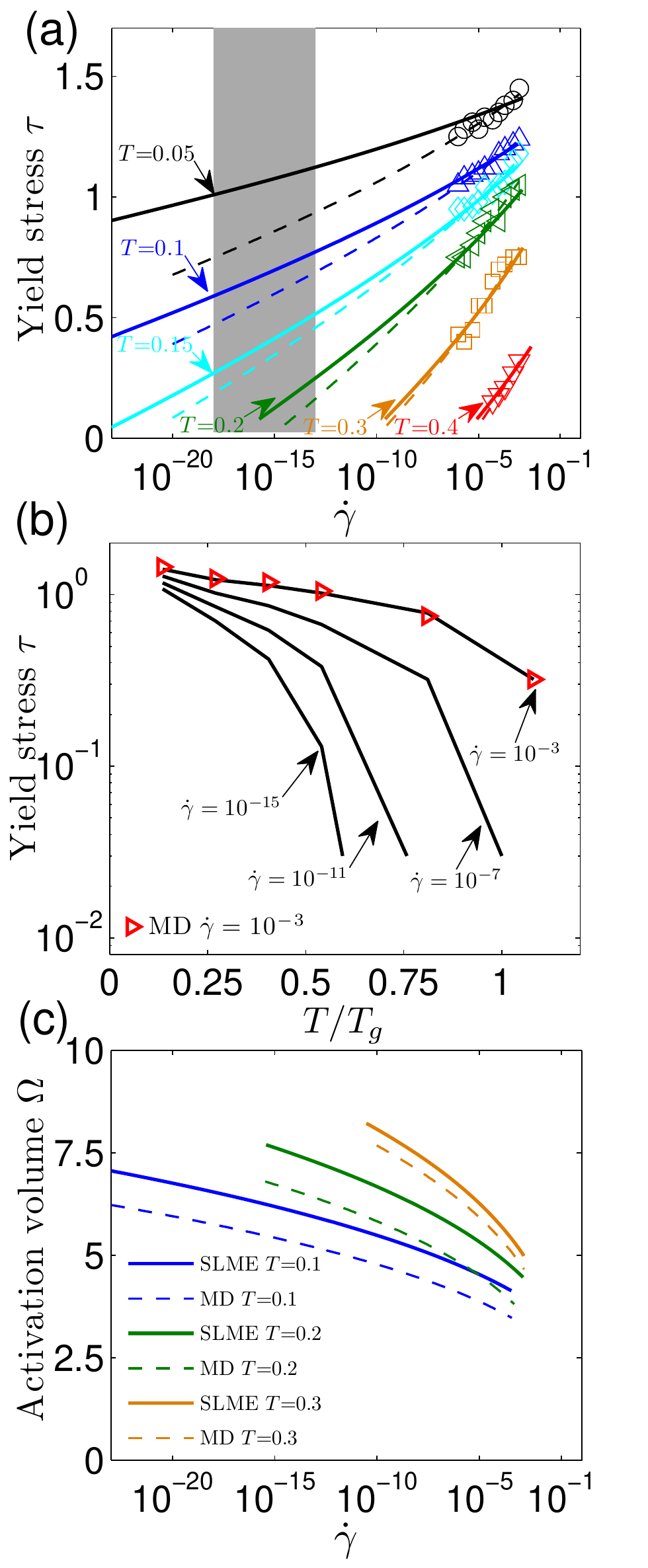} 
\caption{\label{}Comparison of MD (open symbols and their extrapolated dashed lines) and SLME (solid lines) results for (a) yield stress $\tau$ versus strain rate $\dot{\gamma}$, with the experimental strain rate window shaded;   (b) yield stress $\tau$ versus normalized temperature $T/T_{g}$; (c) activation volume $\Omega$ versus strain rate $\dot{\gamma}$.}
\label{fig4} \end{center} \end{figure}

Although qualitative trends similar to Fig. \ref{fig1a} have been observed in recent MD simulations of yielding in amorphous solids~\citep{chengAM2011}, there are notable quantitative discrepancies between the MD and SLME results, especially at low temperature and when high strain rate MD results are extrapolated to slow strain rates.  Explicitly, we follow Eq. (\ref{eq:qatau2}) and plot the yield stress versus strain rate curve for various temperatures in Fig. \ref{fig4}(a), which compares the SLME results (filled symbols and solid lines) to the extrapolation of the MD results (open symbols and dashed lines) to slower strain rates.  What is noticeable is that, while the SLME and MD results agree fairly well for all strain rates for temperatures greater than about $T=0.2$, the difference between the SLME and extrapolated MD results at slower strain rates increases as the temperature decreases. 

These discrepancies at low temperatures are mainly caused by the differing values of $A_{T}$ in Eq. (\ref{eq:qatau2}), where the SLME values of $A_{T}$ are significantly larger than the MD values at low temperatures.  For example, at $T= 0.05=0.14T_{\rm g}$ the SLME $A_T= 9.27$ is nearly two times larger than the MD $A_{T}$=4.86. Physically, this means that while at low temperatures the MD trajectories are dominated by slow dynamics, the SLME trajectories can still explore the entire relevant PES subspace truncated by the given strain rate (or effectively $Q^*$ of Fig. \ref{fig1}) to identify the most favorable free energy basins.  The ability of reaching more stable free-energy metabasins using the SLME approach enables the materials to sustain larger yield stresses.  On the other hand, the MD and SLME predictions for the yield stress are consistent for the MD-accessible strain rate of $\dot{\gamma}=10^{-3}$ for all temperatures.  Not coincidentally, the value of $A_{T}$ for SLME is 7.96 while the $A_{T}$ from MD is 7.56 at $T=0.2$, which reflects the good agreement between MD and SLME as the temperature increases.

\begin{figure} \begin{center} 
\includegraphics[scale=0.32]{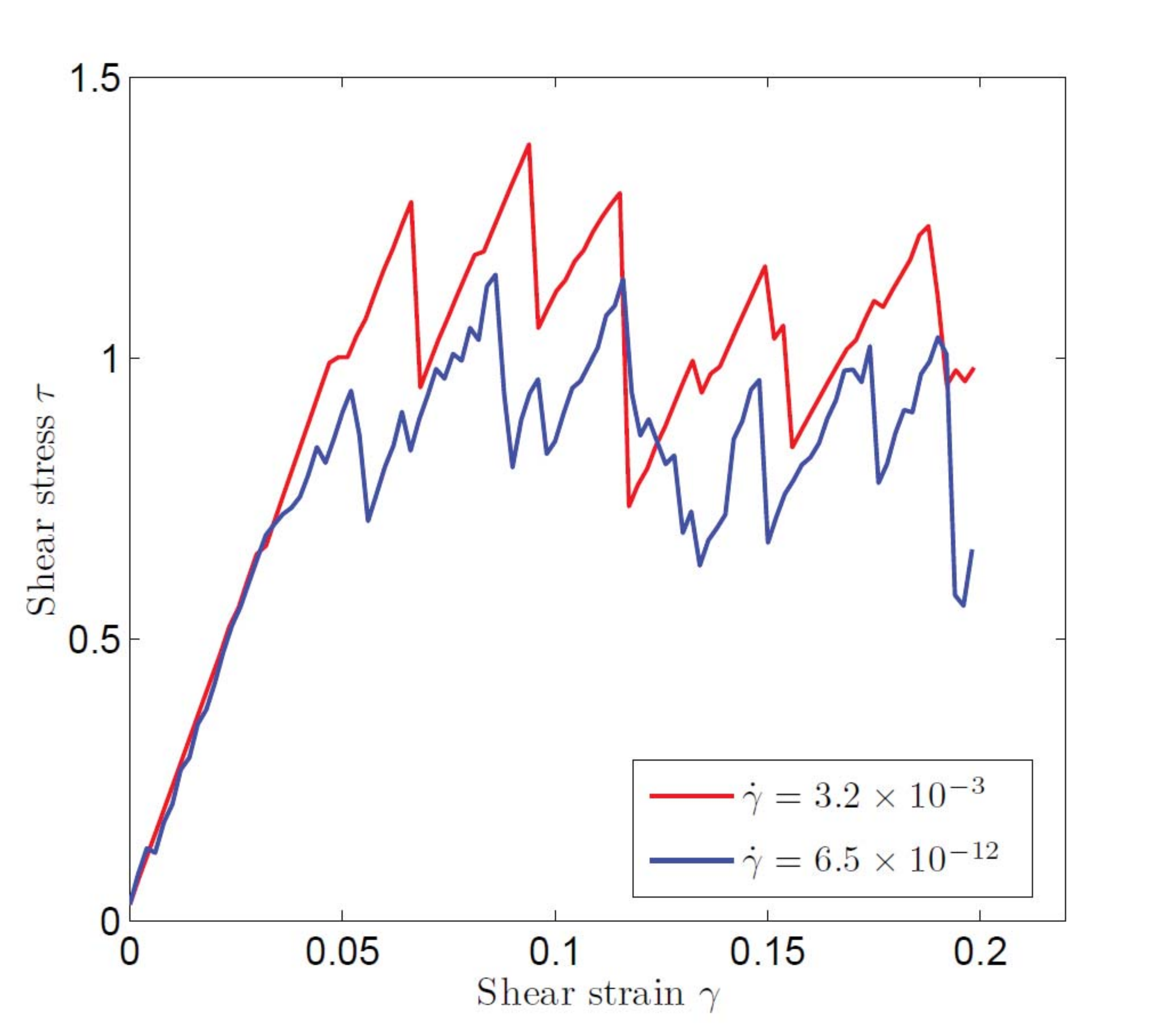}
\caption{\label{}Shear stress $\tau$ versus strain $\gamma$ at $T=0.1$ by SLME for two different strain rates, the first ($\dot{\gamma}=3.2\times10^{-3}$) being comparable to MD, and the second ($\dot{\gamma}=6.5\times10^{-12}$) being experimentally-relevant (i.e. $\approx$10 orders of magnitude smaller than MD).}
\label{fig14} \end{center} \end{figure}

To further quantify the difference introduced due to the discrepancies in $A_{T}$ between MD and SLME at low temperatures, for $T=0.05$ in Fig. \ref{fig4}(a) the yield stress of 1.07 is reached for a laboratory strain rate of $\dot{\gamma}=10^{-15}$, where the laboratory strain rate is about 10 orders of magnitude slower than the typical MD strain rate of $\dot{\gamma}=10^{-5}$.  This yield stress using the MD extrapolation is predicted to occur at a strain rate of $1.5\times10^{-10}$, or a five order of magnitude predicted difference in strain rate.  In conjunction with this, we show in Fig. \ref{fig14} a comparison of the shear stress-strain curve at $T=0.1$ as obtained using the SLME approach, for both an MD strain rate of $3.2\times10^{-3}$ and an experimental one of $6.5\times10^{-12}$.  It can be seen that the major difference at the two strain rates is the reduction in yield stress with decreasing strain rate, while the elastic properties, i.e. the slope of the linear portion of the stress-strain curve, are quite similar.

From Fig. \ref{fig4}(a), and similar to the results to be discussed in Fig. \ref{fig4}(b), it is clear that for as the temperature increases, the agreement between extrapolated MD and SLME improves, even down to very slow strain rates.  However, for lower temperatures, or the temperature range that the mechanical properties of BMGs are most often tested within, there is a significant discrepancy between the extrapolated MD and SLME results at the slower, experimentally-relevant strain rates.  We can quantify this by choosing the representative MD strain rate to be $\dot{\gamma}=10^{-3}$ and the representative experimental strain rate to be $\dot{\gamma}=10^{-15}$.  Taking $\dot{\gamma}=10^{-15}$, we find that for $T=0.05= 0.14 T_{\rm g}$, the extrapolated yield stress using MD would be about $\tau=0.86$, while the yield stress obtained using SLME is about $\tau=1.07$, which is about 26\% larger.  Similarly, the yield stress obtained using SLME for $T=0.10= 0.28T_{\rm g}$ is about 17\% larger than the extrapolated MD value.  

There are three important points we wish to emphasize based on the numerical results shown in Fig. \ref{fig4}. (1) The SLME results agree with the MD results at MD-accessible strain rates for all temperatures.  (2) At slower strain rates, SLME gives quite different results than obtained via directly extrapolating the MD results, particularly for lower temperatures. (3) One needs other independent means, in particular experimental measurements, to assess the accuracy of the SLME results as compared to the extrapolated MD results at the slower, experimentally-relevant strain rates.  However, we note that while the MD and SLME results are obtained from the same PES, only small barriers are crossed using MD due to the well-known strain rate limitations.  Furthermore, the slow strain rate data obtained using the extrapolation of the MD data is based upon the crossings of small barriers at higher strain rates, while the SLME results at slow strain 
rates are based upon climbing over much larger energy barriers. 

\begin{figure*} \begin{center}
\includegraphics[scale=0.2]{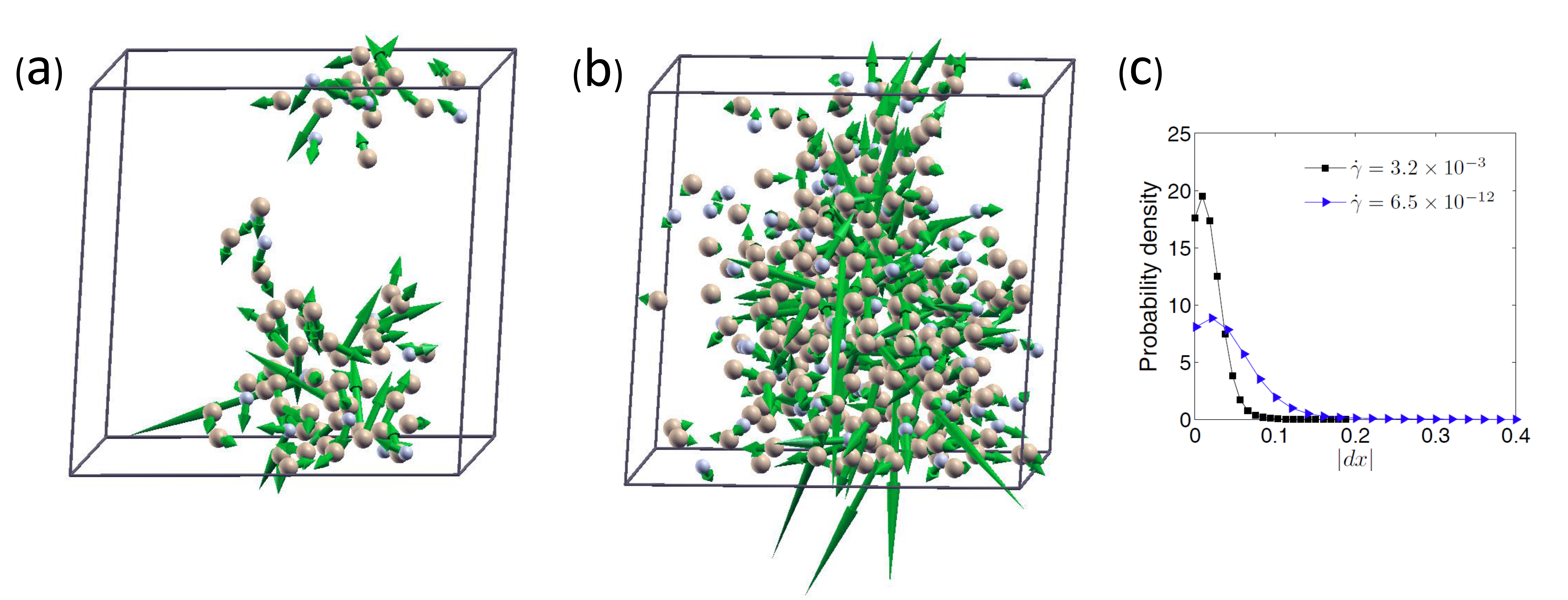} 
\caption{\label{}Snapshot of the number of atoms whose displacement magnitude is larger than 5\% of the maximum displacement magnitude as calculated using the SLME approach at the activated state (i.e. the highest saddle point that is crossed before yield) for strain rates of (a) $\dot{\gamma}=3.2\times10^{-3}$, and (b) $\dot{\gamma}=6.5\times10^{-12}$ and $T=0.1$.  Image (c) shows the probability density of displacements, which demonstrates that at slower strain rates, there are more atoms with larger displacements at yield.  In total, 275 atoms are active by this 5\% criteria in (a) and 335 atoms are active in (b).}
\label{mdabc} \end{center} \end{figure*}

There is another striking result in Fig. \ref{fig4}(b), which is the increasingly strong sensitivity to temperature exhibited by the yield stress for decreasing (slower) strain rates.  In particular, for the experimentally relevant strain rate of $\dot{\gamma}=10^{-15}$, the yield stress decreases nearly two orders of magnitude from very low temperatures to when $T=T_{g}$ is reached.  This dramatic decrease in yield stress occurs because within the SLME framework at very low strain rates, the atomistic system has sufficient time to explore many possible exit paths even out of very deep metabasins~\citep{caoPRE2012,caoPRE2013}. 

We note that a similar result to the deformation map shown in Fig. \ref{fig4}(b) was presented by~\citet{homerAM2009} using a mesoscale finite element model for Vitreloy 1, i.e. a significant decrease in yield strength with increasing temperature at very slow strain rates.  The constitutive assumptions in the~\citet{homerAM2009} model that enables this behavior is that STZs can shear both forward and backward, and that the STZs can interact through their elastic fields.  These assumptions combine, at high temperature, to lead to homogeneous glass flow, with a decrease in strain rate sensitivity with increasing stress.  Interestingly, the same behavior is observed in the SLME results in Fig. \ref{fig4}(b) where no such constitutive assumptions are made, and where the yield stress was obtained solely through PES exploration.  This suggests that the current SLME model is able to capture the experimentally-observed~\citep{schuhAM2007} transition from non-Newtonian flow to Newtonian-flow at temperatures approaching $T_{g}$ and very slow loading rates, which is also correlated with a transition from low to high strain rate sensitivity.
\begin{figure} \begin{center}
\includegraphics[scale=0.66]{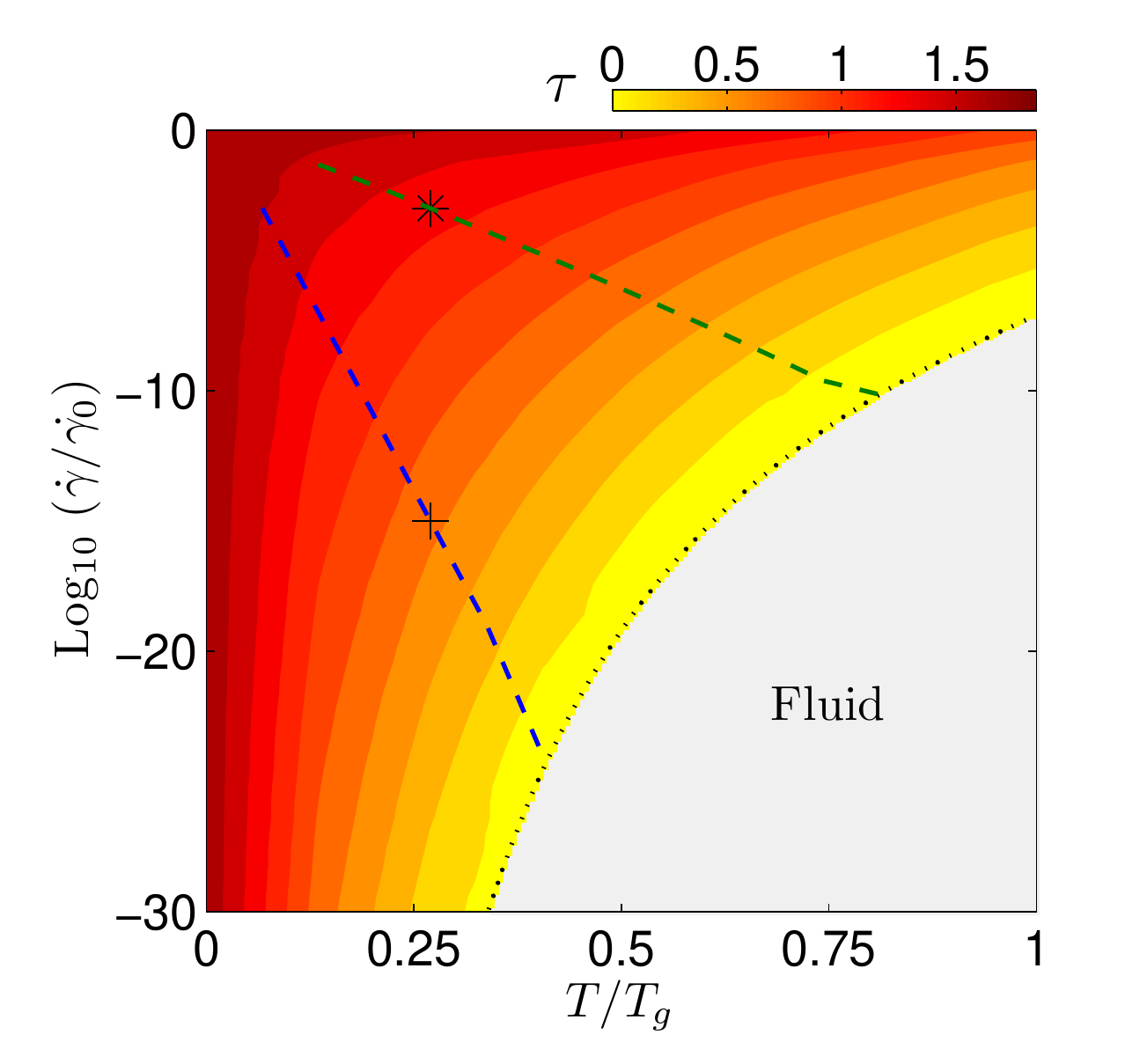} 
\caption{\label{}Contour plot of the shear yield stress $\tau_{y}$ as a function of normalized shear strain rate $\dot{\gamma}/\dot{\gamma_{0}}$ and normalized temperature $T/T_{g}$ computed by the SLME method.  At the experimental strain rate $\dot{\gamma}/\dot{\gamma_0}=10^{-15}$ and room-temperature $T= 0.27T_{\rm g}$ (plus symbol), the dashed line shows that the yield stress is as sensitive to a one order of magnitude decrease in strain rate as it is to a 1.5\%$T_{g}$ increase in temperature.  In contrast, at a typical MD strain rate of $\dot{\gamma}/\dot{\gamma_0}=10^{-3}$ and room-temperature $T= 0.27T_{\rm g}$ (star), the dashed line shows that the yield stress is as sensitive to a one order of magnitude decrease in strain rate as it is to a 5.1\%$T_{g}$ increase in temperature.}
\label{fig3} \end{center} \end{figure}

In addition to quantifying the differences inherent to extrapolating MD simulations of the yield stress to experimentally-relevant strain rates, we also discuss the differences in the calculation of the activation volume $\Omega$.  The activation volume is important because it directly reflects the STZ volume as $\Omega \simeq 0.1V_{\rm STZ}$~\citep{argonAM1979,schuhAM2007}.  The activation volume is defined as the derivative of activation energy with stress, i.e. $\Omega(\dot{\gamma},T) = -\partial{Q^*}/\partial{\tau}$. Because the strain rate $\dot{\gamma}$ and the activation energy $Q^*$ are related by Eq. \ref{eq:tst4}, the activation volume can be written as $\Omega(\dot{\gamma},T) \propto (\partial{\tau}/\partial{\ln{\dot{\gamma}}})^{-1}$, where $\partial{\tau}/\partial{\ln{\dot{\gamma}}}$ is the strain rate sensitivity.

We therefore plot in Fig. \ref{fig4}(c) the activation volume $\Omega$ that is computed using both MD and SLME for three representative temperatures below $T_{g}$ as a function of the shear strain rate $\dot{\gamma}$.  As can be seen, there are again larger differences in the activation volume between the extrapolated MD and SLME predictions for the lower temperatures.  Furthermore, it is clear that for the considered range of temperatures from $T=0.1$ to $T=0.3$ and shear strain rates ranging from $\dot{\gamma}=10^{-3}$ to $10^{-20}$ that the activation volume $\Omega$ as computed using the SLME method generally ranges from 4-8 in reduced LJ units.  Considering the density of the bLJ amorphous solid is around 1.2, this implies that the activation volume $\Omega$ ranges from about 5 to 10 atoms, and that the  activation volume increases with decreasing strain rate $\dot{\gamma}$.  Furthermore, the strain rate sensitivity decreases with decreasing $\dot{\gamma}$, which can be seen in Fig. \ref{fig4}(a), which implies that the yield stress is more sensitive to the strain rate for higher strain rates.

However, for lower temperatures, similar to the observations in Figs. \ref{fig4}(a) and \ref{fig4}(b) for the yield stress, there is a noticeable difference in the activation volume predicted using the extrapolated MD and SLME results at an experimentally-relevant strain rate of $\dot{\gamma}=10^{-15}$.  Specifically, the activation volume for SLME is about 6.3, while the activation volume that is obtained by extrapolating the MD result is about 5.5.  This difference corresponds to a difference of 15\% in the STZ volume estimated by extrapolating the results of high strain rate MD simulations.  Again, this result suggests that caution should be utilized in directly extrapolating MD results down to laboratory strain rates~\citep{chengAM2011}, particularly for the lower temperatures that are the ones of technological interest.

We compare the atomistic motion at yield at the activated state (corresponding to the highest saddle point that is crossed prior to yield) for two different strain rates, $\dot{\gamma}=3.2\times10^{-3}$ and $6.5\times10^{-12}$ in Fig. \ref{mdabc} as obtained using the SLME approach.  What is noteworthy in doing so is that while the number of active atoms is not substantially larger at the slower strain rate, Fig. \ref{mdabc}(c) demonstrates that those atoms typically have larger displacements at yield at slower strain rates.  Furthermore, Fig. \ref{mdabc} demonstrates that, unlike the visualization of defects in crystalline nanostructures using well-defined metrics such as centrosymmetry or common neighbor analysis to identify dislocations, stacking faults and twins, it is difficult to identify precisely the yield mechanism in amorphous solids.  This difficulty is likely the main reason why many researchers prefer to perform two-dimensional studies of yielding in amorphous solids~\citep{falkPRE1998,zinkPRB2006,karmakarPRL2010,maloneyPRE2006,tanguyEPJE2006,tsamadosPRE2009}.

The results of Fig. \ref{fig4} have significant ramifications for the interpretation of previous results reported in the literature.  In particular, as previously noted, a recent work by~\citet{chengAM2011} used the earlier cooperative shear model (CSM) of~\citet{johnsonPRL2005} to extrapolate MD simulation results of Cu$_{64}$Zr$_{36}$ at high strain rates to predict the yield stress of metallic glasses at experimentally-relevant strain rates for a range of temperatures up to $T/T_{g}=300K/787K=0.38$.  Our results suggest that this extrapolation of MD results to lower strain rates via the CSM model may be valid, but only for temperatures larger than about $T/T_{g}=0.15/0.37=0.41$.  Thus, if a similar trend holds for Cu$_{64}$Zr$_{36}$, the yield stresses at experimental strain rates for that material at the temperature range considered by~\citet{chengAM2011} should in fact be considerably ($\sim20-30\%$) higher.

However, the most important implication of Fig. \ref{fig4} is that the SLME simulations and the extrapolation of high strain rate MD simulations follow distinctly different energetic paths during the applied shear deformation at low temperatures and experimental strain rates, which implies that extrapolation of the governing deformation mechanisms from MD strain rates to experimental ones may not be valid. This can be seen by the discrepancy in the yield stress as well as the activation volume between the extrapolated MD and SLME for temperatures less than $T=0.5T_{g}$.

Finally, we plot in Fig. \ref{fig3} the shear yield stress $\tau_{y}$ contour as a function of the normalized strain rate $\dot\gamma/\dot{\gamma_0}$ and normalized temperature $T/T_{\rm g}$, which is directly generated using the SLME method.  Taking as representative a room temperature-like value of $T=0.27T_{\rm g}$ and an experimental strain rate of $\dot{\gamma}/\dot{\gamma_0}=10^{-15}$, we find that $\partial\tau/\partial{\rm log}_{10}(\dot\gamma/\dot{\gamma_0}) = 0.039$ and $\partial\tau/\partial(T/T_{\rm g})= 2.6$.  This establishes the {\it strain rate equivalent of temperature} for yield stress variations.  Namely, the yield stress is found to be as sensitive to a one order of magnitude decrease in strain rate as it is to a $1.5\%T_{\rm g}$ increase in temperature.  This is in contrast to at the MD strain rate of $\dot{\gamma}/\dot{\gamma_0}=10^{-3}$ and room temperature $T=0.27T_{g}$, where the yield stress is found to be as sensitive to a one order of magnitude decrease in strain rate as it is to a $5.1\%T_{\rm g}$ increase in temperature.  

\section{Conclusions}

In conclusion, we have developed a new atomistic computational approach coupling PES exploration techniques and transition state theory to study the strain rate and temperature-dependence of the yield stress for a model bLJ amorphous solid.  This novel approach allows us to study the deformation and failure of amorphous solids at strain rates ranging from MD to experimental, and all relevant temperatures under the melting temperature.  We have found that the results of classical MD simulations can safely be extrapolated down to experimental strain rates for temperatures higher than $0.4T_{g}$.  However, significant differences were found in both the yield stress and the activation volume for lower temperatures between the proposed method and extrapolated MD results, which suggests extrapolation of the governing deformation mechanisms from MD strain rates may not be valid at lower temperatures.  Finally, we have identified the strain rate equivalent of temperature on the yield stress sensitivity at laboratory experimental conditions.  

\section{Acknowledgements} 

All authors acknowledge the support of the NSF through grant CMMI-1234183.  XL also acknowledges the support of the NSF-XSEDE through grant DMR-0900073. All authors acknowledge the comments and suggestions of both reviewers.

\bibliographystyle{model2-names}
\bibliography{biball}

\begin{thebibliography}{39}
\expandafter\ifx\csname natexlab\endcsname\relax\def\natexlab#1{#1}\fi
\expandafter\ifx\csname url\endcsname\relax
  \def\url#1{\texttt{#1}}\fi
\expandafter\ifx\csname urlprefix\endcsname\relax\def\urlprefix{URL }\fi
\providecommand{\eprint}[2][]{\url{#2}}
\providecommand{\bibinfo}[2]{#2}
\ifx\xfnm\relax \def\xfnm[#1]{\unskip,\space#1}\fi
\bibitem[{Argon(1979)}]{argonAM1979}
\bibinfo{author}{Argon, A.S.}, \bibinfo{year}{1979}.
\newblock \bibinfo{title}{Plastic deformation in metallic glasses}.
\newblock \bibinfo{journal}{Acta Metallurgica} \bibinfo{volume}{27},
  \bibinfo{pages}{47--58}.
\bibitem[{Cao et~al.(2009)Cao, Cheng and Ma}]{caoAM2009}
\bibinfo{author}{Cao, A.J.}, \bibinfo{author}{Cheng, Y.Q.},
  \bibinfo{author}{Ma, E.}, \bibinfo{year}{2009}.
\newblock \bibinfo{title}{Structural processes that initiate shear localization
  in metallic glass}.
\newblock \bibinfo{journal}{Acta Materialia} \bibinfo{volume}{57},
  \bibinfo{pages}{5146--5155}.
\bibitem[{Cao et~al.(2012)Cao, Li, Heugle, Park and Lin}]{caoPRE2012}
\bibinfo{author}{Cao, P.}, \bibinfo{author}{Li, M.}, \bibinfo{author}{Heugle,
  R.J.}, \bibinfo{author}{Park, H.S.}, \bibinfo{author}{Lin, X.},
  \bibinfo{year}{2012}.
\newblock \bibinfo{title}{A self-learning metabasin escape algorithm and the
  metabasin correlation lengths of supercooled liquids}.
\newblock \bibinfo{journal}{Physical Review E} \bibinfo{volume}{86},
  \bibinfo{pages}{016710}.
\bibitem[{Cao et~al.(2013)Cao, Park and Lin}]{caoPRE2013}
\bibinfo{author}{Cao, P.}, \bibinfo{author}{Park, H.S.}, \bibinfo{author}{Lin,
  X.}, \bibinfo{year}{2013}.
\newblock \bibinfo{title}{Strain-rate and temperature-driven transition in the
  shear transformation zone for two-dimensional amorphous solids}.
\newblock \bibinfo{journal}{Physical Review E} \bibinfo{volume}{88},
  \bibinfo{pages}{042404}.
\bibitem[{Cheng and Ma(2011a)}]{chengPMS2011}
\bibinfo{author}{Cheng, Y.Q.}, \bibinfo{author}{Ma, E.}, \bibinfo{year}{2011}a.
\newblock \bibinfo{title}{Atomic-level structure and structure-property
  relationship in metallic glasses}.
\newblock \bibinfo{journal}{Progress in Materials Science}
  \bibinfo{volume}{56}, \bibinfo{pages}{379--473}.
\bibitem[{Cheng and Ma(2011b)}]{chengAM2011}
\bibinfo{author}{Cheng, Y.Q.}, \bibinfo{author}{Ma, E.}, \bibinfo{year}{2011}b.
\newblock \bibinfo{title}{Intrinsic shear strength of metallic glass}.
\newblock \bibinfo{journal}{Acta Materialia} \bibinfo{volume}{59},
  \bibinfo{pages}{1800--1807}.
\bibitem[{Dasgupta et~al.(2012)Dasgupta, Hentschel and
  Procaccia}]{dasguptaPRL2012}
\bibinfo{author}{Dasgupta, R.}, \bibinfo{author}{Hentschel, H.G.E.},
  \bibinfo{author}{Procaccia, I.}, \bibinfo{year}{2012}.
\newblock \bibinfo{title}{Microscopic mechanism of shear bands in amorphous
  solids}.
\newblock \bibinfo{journal}{Physical Review Letters} \bibinfo{volume}{109},
  \bibinfo{pages}{255502}.
\bibitem[{Delogu(2008)}]{deloguPRL2008b}
\bibinfo{author}{Delogu, F.}, \bibinfo{year}{2008}.
\newblock \bibinfo{title}{Atomic mobility and strain localization in amorphous
  metals}.
\newblock \bibinfo{journal}{Physical Review Letters} \bibinfo{volume}{100},
  \bibinfo{pages}{075901}.
\bibitem[{Falk and Langer(1998)}]{falkPRE1998}
\bibinfo{author}{Falk, M.L.}, \bibinfo{author}{Langer, J.S.},
  \bibinfo{year}{1998}.
\newblock \bibinfo{title}{Dynamics of viscoplastic deformation in amorphous
  solids}.
\newblock \bibinfo{journal}{Physical Review E} \bibinfo{volume}{57},
  \bibinfo{pages}{7192--7205}.
\bibitem[{Henkelman and Jonsson(1999)}]{henkelmanJCP1999}
\bibinfo{author}{Henkelman, G.}, \bibinfo{author}{Jonsson, H.},
  \bibinfo{year}{1999}.
\newblock \bibinfo{title}{Dimer method for finding saddle points on high
  dimensional potential surfaces using only first derivatives}.
\newblock \bibinfo{journal}{Journal of Computational Physics}
  \bibinfo{volume}{111}, \bibinfo{pages}{7010--7022}.
\bibitem[{Homer and Schuh(2009)}]{homerAM2009}
\bibinfo{author}{Homer, E.R.}, \bibinfo{author}{Schuh, C.A.},
  \bibinfo{year}{2009}.
\newblock \bibinfo{title}{Mesoscale modeling of amorphous metals by shear
  transformation zone dynamics}.
\newblock \bibinfo{journal}{Acta Materialia} \bibinfo{volume}{57},
  \bibinfo{pages}{2823--2833}.
\bibitem[{Johnson and Samwer(2005)}]{johnsonPRL2005}
\bibinfo{author}{Johnson, W.L.}, \bibinfo{author}{Samwer, K.},
  \bibinfo{year}{2005}.
\newblock \bibinfo{title}{A universal criterion for plastic yielding of
  metallic glasses with a $(t/t_{g})^{2/3}$ temperature dependence}.
\newblock \bibinfo{journal}{Physical Review Letters} \bibinfo{volume}{95},
  \bibinfo{pages}{195501}.
\bibitem[{Karmakar et~al.(2010)Karmakar, Lemaitre, Lerner and
  Procaccia}]{karmakarPRL2010}
\bibinfo{author}{Karmakar, S.}, \bibinfo{author}{Lemaitre, A.},
  \bibinfo{author}{Lerner, E.}, \bibinfo{author}{Procaccia, I.},
  \bibinfo{year}{2010}.
\newblock \bibinfo{title}{Predicting plastic flow events in athermal
  shear-strained amorphous solids}.
\newblock \bibinfo{journal}{Physical Review Letters} \bibinfo{volume}{104},
  \bibinfo{pages}{215502}.
\bibitem[{Kob and Andersen(1995)}]{kobPRE1995}
\bibinfo{author}{Kob, W.}, \bibinfo{author}{Andersen, H.C.},
  \bibinfo{year}{1995}.
\newblock \bibinfo{title}{Testing mode-coupling theory for a supercooled binary
  lennard-jones mixture: the van hove correlation function}.
\newblock \bibinfo{journal}{Physical Review E} \bibinfo{volume}{51},
  \bibinfo{pages}{4626--4641}.
\bibitem[{Koziatek et~al.(2013)Koziatek, Barrat, Derlet and
  Rodney}]{koziatekPRB2013}
\bibinfo{author}{Koziatek, P.}, \bibinfo{author}{Barrat, J.L.},
  \bibinfo{author}{Derlet, P.}, \bibinfo{author}{Rodney, D.},
  \bibinfo{year}{2013}.
\newblock \bibinfo{title}{Inverse meyer-neldel behavior for activated processes
  in model glasses}.
\newblock \bibinfo{journal}{Physical Review B} \bibinfo{volume}{87},
  \bibinfo{pages}{224105}.
\bibitem[{Kushima et~al.(2009a)Kushima, Lin, Li, Eapen, Mauro, Qian, Diep and
  Yip}]{kushimaJCP2009}
\bibinfo{author}{Kushima, A.}, \bibinfo{author}{Lin, X.}, \bibinfo{author}{Li,
  J.}, \bibinfo{author}{Eapen, J.}, \bibinfo{author}{Mauro, J.C.},
  \bibinfo{author}{Qian, X.}, \bibinfo{author}{Diep, P.}, \bibinfo{author}{Yip,
  S.}, \bibinfo{year}{2009}a.
\newblock \bibinfo{title}{Computing the viscosity of supercooled liquids}.
\newblock \bibinfo{journal}{Journal of Chemical Physics} \bibinfo{volume}{130},
  \bibinfo{pages}{224504}.
\bibitem[{Kushima et~al.(2009b)Kushima, Lin, Li, Qian, Eapen, Mauro, Diep and
  Yip}]{kushimaJCP2009_2}
\bibinfo{author}{Kushima, A.}, \bibinfo{author}{Lin, X.}, \bibinfo{author}{Li,
  J.}, \bibinfo{author}{Qian, X.}, \bibinfo{author}{Eapen, J.},
  \bibinfo{author}{Mauro, J.}, \bibinfo{author}{Diep, P.},
  \bibinfo{author}{Yip, S.}, \bibinfo{year}{2009}b.
\newblock \bibinfo{title}{Computing the viscosity of supercooled liquids. ii.
  silica and strong-fragile crossover behavior}.
\newblock \bibinfo{journal}{The Journal of chemical physics}
  \bibinfo{volume}{131}, \bibinfo{pages}{164505}.
\bibitem[{Laio and Parrinello(2002)}]{laioPNAS2002}
\bibinfo{author}{Laio, A.}, \bibinfo{author}{Parrinello, M.},
  \bibinfo{year}{2002}.
\newblock \bibinfo{title}{Escaping free-energy minima}.
\newblock \bibinfo{journal}{Proceedings of the National Academy of Science}
  \bibinfo{volume}{99}, \bibinfo{pages}{12562--12566}.
\bibitem[{Li et~al.(2011)Li, Kushima, Eapen, Lin, Qian, Mauro, Diep and
  Yip}]{liPO2011}
\bibinfo{author}{Li, J.}, \bibinfo{author}{Kushima, A.},
  \bibinfo{author}{Eapen, J.}, \bibinfo{author}{Lin, X.},
  \bibinfo{author}{Qian, X.}, \bibinfo{author}{Mauro, J.},
  \bibinfo{author}{Diep, P.}, \bibinfo{author}{Yip, S.}, \bibinfo{year}{2011}.
\newblock \bibinfo{title}{Computing the viscosity of supercooled liquids:
  Markov network model}.
\newblock \bibinfo{journal}{PLoS One} \bibinfo{volume}{6},
  \bibinfo{pages}{e17909}.
\bibitem[{Maloney and Lemaitre(2006)}]{maloneyPRE2006}
\bibinfo{author}{Maloney, C.E.}, \bibinfo{author}{Lemaitre, A.},
  \bibinfo{year}{2006}.
\newblock \bibinfo{title}{Amorphous systems in athermal, quasistatic shear}.
\newblock \bibinfo{journal}{Physical Review E} \bibinfo{volume}{74},
  \bibinfo{pages}{016118}.
\bibitem[{Mayr(2006)}]{mayrPRL2006}
\bibinfo{author}{Mayr, S.G.}, \bibinfo{year}{2006}.
\newblock \bibinfo{title}{Activation energy of shear transformation zones: a
  key for understanding rheology of glasses and liquids}.
\newblock \bibinfo{journal}{Physical Review Letters} \bibinfo{volume}{97},
  \bibinfo{pages}{195501}.
\bibitem[{Miron and Fichthorn(2004)}]{mironPRL2004}
\bibinfo{author}{Miron, R.A.}, \bibinfo{author}{Fichthorn, K.A.},
  \bibinfo{year}{2004}.
\newblock \bibinfo{title}{Multiple-time scale accelerated molecular dynamics:
  addressing the small-barrier problem}.
\newblock \bibinfo{journal}{Physical Review Letters} \bibinfo{volume}{93},
  \bibinfo{pages}{128301}.
\bibitem[{Mirona and Fichthorn(2003)}]{mirona03}
\bibinfo{author}{Mirona, R.A.}, \bibinfo{author}{Fichthorn, K.A.},
  \bibinfo{year}{2003}.
\newblock \bibinfo{title}{Computing the viscosity of supercooled liquids. ii.
  silica and strong-fragile crossover behavior}.
\newblock \bibinfo{journal}{The Journal of chemical physics}
  \bibinfo{volume}{119}, \bibinfo{pages}{6210}.
\bibitem[{Murali et~al.(2011)Murali, Guo, Zhang, Narasimhan, Li and
  Gao}]{muraliPRL2011}
\bibinfo{author}{Murali, P.}, \bibinfo{author}{Guo, T.F.},
  \bibinfo{author}{Zhang, Y.W.}, \bibinfo{author}{Narasimhan, R.},
  \bibinfo{author}{Li, Y.}, \bibinfo{author}{Gao, H.J.}, \bibinfo{year}{2011}.
\newblock \bibinfo{title}{Atomic scale fluctuations govern brittle fracture and
  cavitation behavior in metallic glasses}.
\newblock \bibinfo{journal}{Physical Review Letters} \bibinfo{volume}{107},
  \bibinfo{pages}{215501}.
\bibitem[{Park et~al.(2006)Park, Gall and Zimmerman}]{parkJMPS2006}
\bibinfo{author}{Park, H.S.}, \bibinfo{author}{Gall, K.},
  \bibinfo{author}{Zimmerman, J.A.}, \bibinfo{year}{2006}.
\newblock \bibinfo{title}{Deformation of {FCC} nanowires by twinning and slip}.
\newblock \bibinfo{journal}{Journal of the Mechanics and Physics of Solids}
  \bibinfo{volume}{54}, \bibinfo{pages}{1862--1881}.
\bibitem[{Rodney and Schuh(2009a)}]{rodneyPRL2009}
\bibinfo{author}{Rodney, D.}, \bibinfo{author}{Schuh, C.A.},
  \bibinfo{year}{2009}a.
\newblock \bibinfo{title}{Distribution of thermally activated plastic events in
  a flowing glass}.
\newblock \bibinfo{journal}{Physical Review Letters} \bibinfo{volume}{102},
  \bibinfo{pages}{235503}.
\bibitem[{Rodney and Schuh(2009b)}]{rodneyPRB2009}
\bibinfo{author}{Rodney, D.}, \bibinfo{author}{Schuh, C.A.},
  \bibinfo{year}{2009}b.
\newblock \bibinfo{title}{Yield stress in metallic glasses: the
  jamming-unjamming transition studied through monte carlo simulations based on
  the activation-relaxation technique}.
\newblock \bibinfo{journal}{Physical Review B} \bibinfo{volume}{80},
  \bibinfo{pages}{184203}.
\bibitem[{Rodney et~al.(2011)Rodney, Tanguy and Vandembroucq}]{rodneyMSMSE2011}
\bibinfo{author}{Rodney, D.}, \bibinfo{author}{Tanguy, A.},
  \bibinfo{author}{Vandembroucq, D.}, \bibinfo{year}{2011}.
\newblock \bibinfo{title}{Modeling the mechanics of amorphous solids at
  different length and time scale}.
\newblock \bibinfo{journal}{Modelling and Simulation in Materials Science and
  Engineering} \bibinfo{volume}{19}, \bibinfo{pages}{083001}.
\bibitem[{Ryu et~al.(2011)Ryu, Kang and Cai}]{ryuPNAS2011}
\bibinfo{author}{Ryu, S.}, \bibinfo{author}{Kang, K.}, \bibinfo{author}{Cai,
  W.}, \bibinfo{year}{2011}.
\newblock \bibinfo{title}{Entropic effect on the rate of dislocation
  nucleation}.
\newblock \bibinfo{journal}{Proceedings of the National Academy of Science}
  \bibinfo{volume}{108}, \bibinfo{pages}{5174--5178}.
\bibitem[{Sastry et~al.(1998)Sastry, Debenedetti and
  Stillinger}]{sastryNATURE1998}
\bibinfo{author}{Sastry, S.}, \bibinfo{author}{Debenedetti, P.G.},
  \bibinfo{author}{Stillinger, F.H.}, \bibinfo{year}{1998}.
\newblock \bibinfo{title}{Signatures of distinct dynamical regimes in the
  energy landscape of a glass-forming liquic}.
\newblock \bibinfo{journal}{Nature} \bibinfo{volume}{393},
  \bibinfo{pages}{554--557}.
\bibitem[{Schuh et~al.(2007)Schuh, Hufnagel and Ramamurty}]{schuhAM2007}
\bibinfo{author}{Schuh, C.A.}, \bibinfo{author}{Hufnagel, T.C.},
  \bibinfo{author}{Ramamurty, U.}, \bibinfo{year}{2007}.
\newblock \bibinfo{title}{Mechanical behavior of amorphous alloys}.
\newblock \bibinfo{journal}{Acta Materialia} \bibinfo{volume}{55},
  \bibinfo{pages}{4067--4109}.
\bibitem[{Shi and Falk(2007)}]{shiAM2007}
\bibinfo{author}{Shi, Y.}, \bibinfo{author}{Falk, M.L.}, \bibinfo{year}{2007}.
\newblock \bibinfo{title}{Stress-induced structural transformation and shear
  banding during simulated nanoindentation of a metallic glass}.
\newblock \bibinfo{journal}{Acta Materialia} \bibinfo{volume}{55},
  \bibinfo{pages}{4317--4324}.
\bibitem[{Shimizu et~al.(2006)Shimizu, Ogata and Li}]{shimizuAM2006}
\bibinfo{author}{Shimizu, F.}, \bibinfo{author}{Ogata, S.},
  \bibinfo{author}{Li, J.}, \bibinfo{year}{2006}.
\newblock \bibinfo{title}{Yield point of metallic glass}.
\newblock \bibinfo{journal}{Acta Materialia} \bibinfo{volume}{54},
  \bibinfo{pages}{4293--4298}.
\bibitem[{Tanguy et~al.(2006)Tanguy, Leonforte and Barrat}]{tanguyEPJE2006}
\bibinfo{author}{Tanguy, A.}, \bibinfo{author}{Leonforte, F.},
  \bibinfo{author}{Barrat, J.L.}, \bibinfo{year}{2006}.
\newblock \bibinfo{title}{Plastic response of a 2d lennard-jones amorphous
  solid: detailed analysis of the local rearrangements at very slow strain
  rate}.
\newblock \bibinfo{journal}{The European Physical Journal E}
  \bibinfo{volume}{20}, \bibinfo{pages}{355--364}.
\bibitem[{Tsamados et~al.(2009)Tsamados, Tanguy, Goldenberg and
  Barrat}]{tsamadosPRE2009}
\bibinfo{author}{Tsamados, M.}, \bibinfo{author}{Tanguy, A.},
  \bibinfo{author}{Goldenberg, C.}, \bibinfo{author}{Barrat, J.L.},
  \bibinfo{year}{2009}.
\newblock \bibinfo{title}{Local elasticity map and plasticity in a model
  lennard-jones solid}.
\newblock \bibinfo{journal}{Physical Review E} \bibinfo{volume}{80},
  \bibinfo{pages}{026112}.
\bibitem[{Voter(1997)}]{voterPRL1997}
\bibinfo{author}{Voter, A.F.}, \bibinfo{year}{1997}.
\newblock \bibinfo{title}{Hyperdynamics: accelerated molecular dynamics of
  infrequent events}.
\newblock \bibinfo{journal}{Physical Review Letters} \bibinfo{volume}{78},
  \bibinfo{pages}{3908--3911}.
\bibitem[{Wales(2003)}]{wales2003}
\bibinfo{author}{Wales, D.J.}, \bibinfo{year}{2003}.
\newblock \bibinfo{title}{Energy landscapes: With applications to clusters,
  biomolecules and glasses}.
\newblock \bibinfo{publisher}{Cambridge University Press}.
\bibitem[{Zhu et~al.(2008)Zhu, Li, Samanta, Leach and Gall}]{zhuPRL2008}
\bibinfo{author}{Zhu, T.}, \bibinfo{author}{Li, J.}, \bibinfo{author}{Samanta,
  A.}, \bibinfo{author}{Leach, A.}, \bibinfo{author}{Gall, K.},
  \bibinfo{year}{2008}.
\newblock \bibinfo{title}{Temperature and strain-rate dependence of surface
  dislocation nucleation}.
\newblock \bibinfo{journal}{Physical Review Letters} \bibinfo{volume}{100},
  \bibinfo{pages}{025502}.
\bibitem[{Zink et~al.(2006)Zink, Samwer, Johnson and Mayr}]{zinkPRB2006}
\bibinfo{author}{Zink, M.}, \bibinfo{author}{Samwer, K.},
  \bibinfo{author}{Johnson, W.L.}, \bibinfo{author}{Mayr, S.G.},
  \bibinfo{year}{2006}.
\newblock \bibinfo{title}{Plastic deformation of metallic glasses: size of
  shear transformation zones from molecular dynamics simulations}.
\newblock \bibinfo{journal}{Physical Review B} \bibinfo{volume}{73},
  \bibinfo{pages}{172203}.

\end{thebibliography}

\end{document}